\documentclass[12pt]{iopart}
\usepackage{hyperref}
\usepackage{graphicx}
\usepackage{float}
\usepackage{caption}
\usepackage{longtable, multirow}
\usepackage[shortlabels]{enumitem}

\begin{document}

\title[GW anomaly detection using Temporal Outlier Factor]{\texttt{UniMAP}: Model-free detection of unclassified noise transients in LIGO-Virgo data using the Temporal Outlier Factor}

\author{J. Ding$^{1,2}$, R. Ng$^2$ and J. McIver$^1$}
\address{$^1$ Department of Physics and Astronomy, University of British Columbia, Vancouver, British Columbia, V6T1Z4, Canada}
\address{$^2$ Department of Computer Science, University of British Columbia, Vancouver, British Columbia, V6T1Z4, Canada}
\ead{\mailto{julianzding@alumni.ubc.ca}, \mailto{rng@cs.ubc.ca}, \mailto{mciver@phas.ubc.ca}}
\vspace{10pt}
\begin{indented}
\item[]September 2021
\end{indented}

\begin{abstract}
Data from current gravitational wave detectors contains a high rate of transient noise (glitches) that can trigger false detections and obscure true astrophysical events. Existing noise-detection algorithms largely rely on model-based methods that may miss noise transients unwitnessed by auxiliary sensors or with exotic morphologies. We propose the Unicorn Multi-window Anomaly-detection Pipeline (UniMAP): a model-free algorithm to identify and characterize transient noise leveraging the Temporal Outlier Factor (TOF) via a multi-window data-resampling scheme. We show this windowing scheme extends the anomaly detection capabilities of the TOF algorithm to resolve noise transients of arbitrary morphology and duration. We demonstrate the efficacy of this pipeline in detecting glitches during LIGO and Virgo's third observing run, and discuss potential applications.
\end{abstract}
\vspace{2pc}

\submitto{\CQG}

\section{Introduction}

The kilometer-scale interferometers Advanced Laser Interferometer Gravitational-Wave Observatory (LIGO) \cite{aligo} and Advanced Virgo \cite{virgo} are able to detect distance changes between their mirrors of up to $10^{-5}$ the width of a proton. As such, these interferometers are some of the most sensitive instruments ever constructed. To date, they have confidently detected gravitational waves from over 100 compact binary coalescences \cite{gwtc1}\cite{gwtc2}\cite{gwtc3}. However, accompanying such unprecedented sensitivity is a complex challenge of noise filtering for both extracting astrophysical signals of interest from the noise background and preventing noise transients from being misidentified as astrophysical events.

The raw strain data produced by gravitational wave detectors contains persistent and transient noise features \cite{detchar}. Non-Gaussian noise transients can mask or mimic true transient astrophysical signals \cite{davis2020}. These noise transients, also known as \textit{glitches}, likely arise from conditions within the instruments or the surrounding environment \cite{noise_sources}; however, the precise sources of many are unknown \cite{detchar}. Currently, the standard method for characterizing noise transients in LIGO and Virgo data is to first identify them with an excess power detection algorithm, such as Omicron \cite{omicron} or SNAX \cite{snax}, and then classify them by shared sources or morphologies \cite{detchar}. There are a variety of classification approaches in the literature, including algorithms that use information from auxiliary channels that measure the behavior of the detectors and their environment, such as hveto~\cite{Smith2011} and iDQ~\cite{Essick2021}. Complementary approaches classify glitches based on their time-frequency morphology in GW detector data, including GravitySpy \cite{gspy} and other morphology-based classifiers using images~\cite{Razzano2018} and wavelets~\cite{Cuoco2018}. iDQ~\cite{Godwin2020} and other tools have been included as part of LIGO-Virgo candidate event validation procedures, both in near-real time in response to public alerts~\cite{lowlatency} and in higher latency to vet GW events for catalogs~\cite{detchar}. For example,  iDQ, hveto, and GravitySpy were employed in LIGO-Virgo event validation during the most recent observing run~\cite{detchar}.

Unfortunately, previously identified glitch classes do not exhaustively span all possible glitch morphologies, and many exotic glitch events are not witnessed by auxiliary sensors or fall outside of Gravity Spy's existing training set. Additionally, long noise transients on the order of greater than 100 seconds can elude detection by wavelet-based excess power algorithms such as Omicron. These poorly resolved glitches can then trigger false positives in the gravitational wave search pipeline, such as in the case of retracted event S190518bb \cite{eq_gcn} during the third observing run. S190518bb was indirectly classified as a likely binary neutron star merger when, in actuality, it was triggered by transient noise due to an earthquake at the LIGO-Hanford detector \cite{eq_retraction}. Therefore, to improve data quality pipeline sensitivity to these unanticipated noise transients, a model-free anomaly detection method requiring no prior understanding of glitch source or morphology is desirable.

\section{Temporal Outlier Factor}

The problem of algorithmically detecting anomalies in time-series data arises in many applications within and outside gravitational wave astronomy. With the rise of data mining and machine learning as powerful research tools in recent decades, reliable methods for automating the removal of outliers in large datasets are in high demand. Time series data, in particular, presents the additional challenge of being inherently multi-dimensional, but having no geometric relationship between the time dimension and amplitude dimension(s). Indeed, before we can begin detecting anomalies in any time series, we must first choose a definition for what constitutes an anomaly in the first place.

To detect noise transients in interferometer data, we require an anomaly detection algorithm that can pick out \textit{subsequences} of data in a time series. Therefore, we exclude methods such as distance-based outliers \cite{distancebasedoutlier} and conformal $k$-nearest-neighbor anomaly detection schemes \cite{conformalknn} that focus on detecting individual outlying data points. In particular, we also exclude adaptations of the Local Outlier Factor \cite{lof} for univariate time series; in practice, the anomalous data subsequence often forms its own cluster and is therefore largely undetectable except for points making up the rising and falling edges. As previously stated, we also require the anomaly detection algorithm to be model-free. In addition to classifiers like Gravity Spy that require prior knowledge of glitch morphology, this also excludes methods such as sparse dictionary data representations \cite{carrera2016ecg} that require past non-outlier subsequences as a template against which to compare new data.

Of the remaining eligible algorithms, the most popular is the time series discord \cite{hotsax}, which performs a search over all subsequences of a specified length within the data of interest to find one that is maximally different from the rest. Unfortunately, the brute-force implementation of time series discord has an untenable time complexity of $O(n^2)$ on data of length $n$ (though the use of HOT-SAX can improve runtime by a $\sim 10^3$ constant factor \cite{hotsax}). A more recent anomaly detection algorithm approaches the problem from a promising alternative angle: the Temporal Outlier Factor (TOF) \cite{tof}.

The definition of a time series anomaly provided by TOF is based on two observations:
\begin{enumerate}
    \item Anomalies in time series are \textit{unique}---they are aperiodic, and the data points making up an anomaly tend to be close to each other in time.
    \item Anomalies in time series are \textit{distinct}---data trends within an anomaly differ significantly in morphology from the surrounding trends, often in a visually distinct manner.
\end{enumerate}
Observation (i) can be made quantitative via temporal clustering: given a set of data points that are suspected to be part of an anomaly, we can measure the average temporal separation between them. Observation (ii) is defined in much looser terms, though not by choice; quantitatively characterizing the distinctness of a particular trend in the context of a larger stretch of data is a problem with no apparent algorithmic analog. TOF, however, offers an innovative approach by leveraging the time-delay embedding procedure described in Takens's Embedding Theorem \cite{Takens}.

Given a one-dimensional time series $x(t)$, one can embed the data into a $d$-dimensional space by assigning new spatial coordinates using some time delay $\tau$:
\begin{equation}\label{eq:tof_embedding}
    \mathbf{x}(t) = \left(\underbrace{x(t),\ x(t+\tau), \ \dots, \ x\left(t+(d-1)\tau\right)}_d\right)
\end{equation}
\begin{figure}[t]
    \centering
    \includegraphics[scale=0.38]{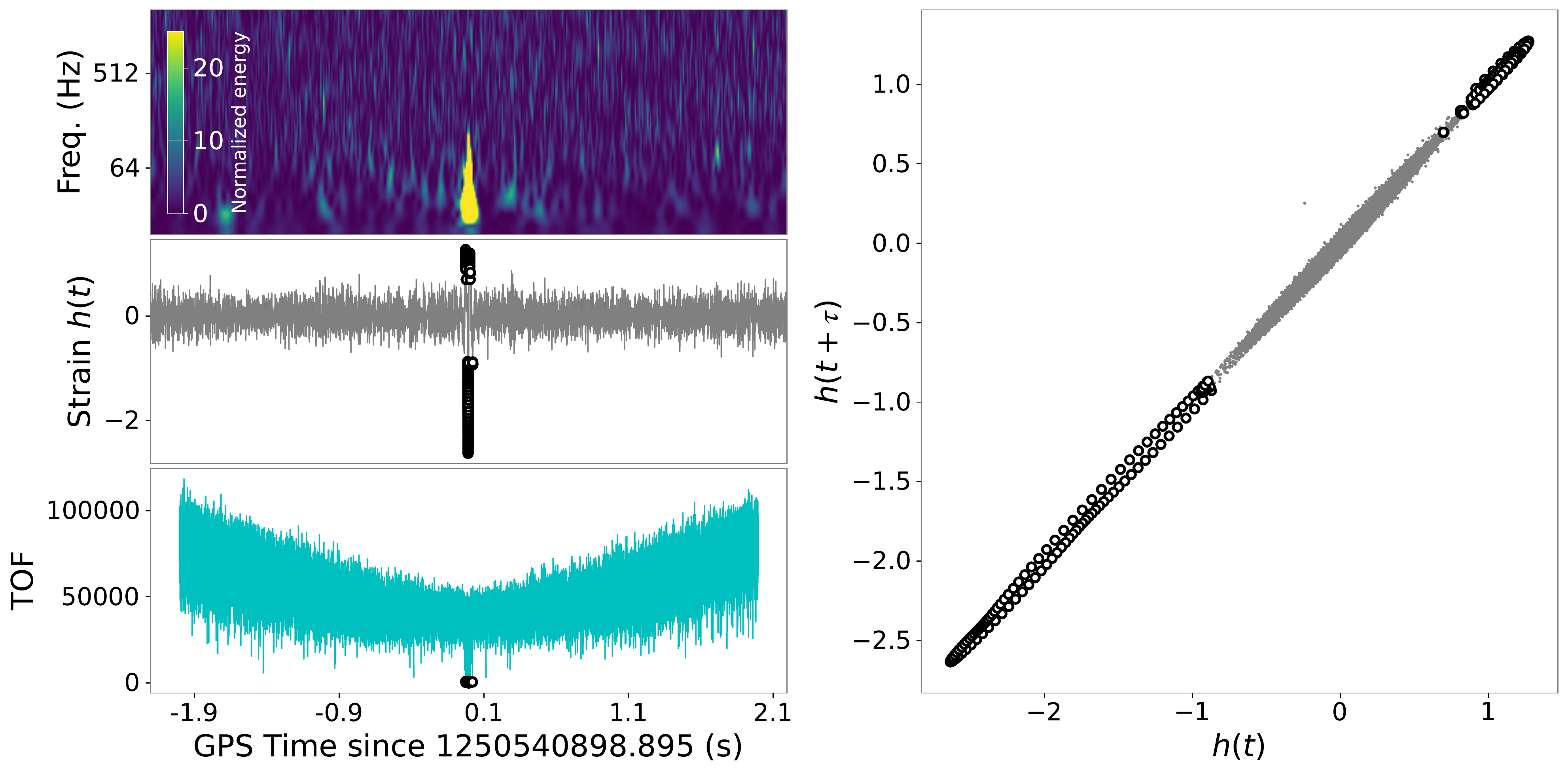}
    \caption{Visual demonstration of the Temporal Outlier Factor anomaly detection algorithm. The top two plots on the left are an example of a \textit{Tomte} glitch in gravitational wave interferometer strain data at a sampling rate of 4192 Hz---the top plot is a Q-transform spectrogram (in the style of those used by the Gravity Spy image classifier~\cite{chatterji2004}), and the middle plot is the raw time series. The corresponding TOF scores for each point in the time series are shown in the bottom left plot, with circles corresponding to points with TOFs below the detection threshold. The right plot shows the corresponding two-dimensional embedded path generated via the time-delay embedding procedure using a delay of $\tau=1$ and embedding dimension of $d=2$. The TOF calculations were performed with a spatial clustering of $k=9$ nearest neighbors and a maximum detectable event length of $M=1024$.}
    \label{fig:2d_embed}
\end{figure}
In a very loose interpretation of Takens's theorem, the embedded time series $\mathbf{x}(t)$ has the property that nearby points belong to morphologically similar data trends\footnote{To be more precise, Takens's Embedding Theorem states that the embedded path $\mathbf{x}(t)$ is topologically equivalent to the specific path through the phase space of the dynamical system that generated the observed time series $x(t)$ if the embedding dimension is sufficiently high, i.e. $d > 2n$ for an $n$-dimensional dynamical system. In practice, however, the dimensionality of the generating dynamical system for an observed time series is often unknown, especially in the case of exotic noise transients.}. Thus, we can expect data trends that differ significantly from the norm to be clustered elsewhere in the embedded space than non-anomalous data (visually demonstrated in Figure \ref{fig:2d_embed}).

We now arrive at the definition of a \textit{unicorn}: a data trend with constituent data points close together in time and corresponding spatially clustered points in the time-delay embedded space. The Temporal Outlier Factor of the data point at time $t$ in a time series $x$ is therefore defined as 
\begin{equation}\label{eq:tof}
    \texttt{TOF}[x,\ t] = \sqrt{\frac{1}{k}\sum_{i\in kNN}\left(t - t_i\right)^2}
\end{equation}
where $k$NN are the $k$ nearest neighbors to the data point of interest in the time-delay embedded path $\mathbf{x}$, and $t_i$ is the corresponding temporal position of each of these $k$ data points in the original time series. We used Euclidean distance as the distance metric between data points in the embedded space.

TOF is a number that describes the normalcy of a particular data point in a time series using the unicorn definition of anomalies---a low TOF score corresponds to a data point that is part of a unicorn. The TOF threshold $\theta$ below which a data point is considered anomalous is set as
\begin{equation}\label{eq:tof_thresh}
    \theta = \sqrt{\frac{1}{k}\sum_{i=0}^{k-1}(M-i)^2}
\end{equation}
where $M$ is the longest detectable unicorn. Note that the derivations shown in \cite{tof} include a parameter $\Delta t$ for the time spacing between individual data points, but for our formulation, we define $\Delta t \equiv 1$ and consider the parameters $\tau$ and $M$ purely in units of (number of data points) for notational simplicity.

The TOF anomaly detection algorithm therefore defines an anomaly $\mathbf{a}$ in a time series $x(t)$ as all the times $t^*$ with corresponding TOF values below the threshold $\theta$:
\begin{equation}\label{eq:tof_anomaly}
    \mathbf{a} = \{t^* \ \forall \ \texttt{TOF}[x,\ t^*] < \theta\}
\end{equation}

\section{Unicorn Multi-window Anomaly-detection Pipeline}

The unicorn definition provides an entirely \textit{a priori} description of time series anomalies and is an excellent candidate for a model-free anomaly detection procedure. However, the TOF algorithm contains several adjustable parameters that must be fine-tuned:
\begin{itemize}
    \item $\tau$, the time delay used in the time-delay embedding procedure
    \item $d$, the embedding dimension used in the time-delay embedding procedure
    \item $k$, the number of nearest neighbors used during spatial clustering
    \item $M$, the maximum detectable anomaly length
\end{itemize}
For anomalies of different lengths, we can use machine learning to optimize the configuration of these parameters.

\begin{figure}[t]
    \centering
    \includegraphics[scale=0.3]{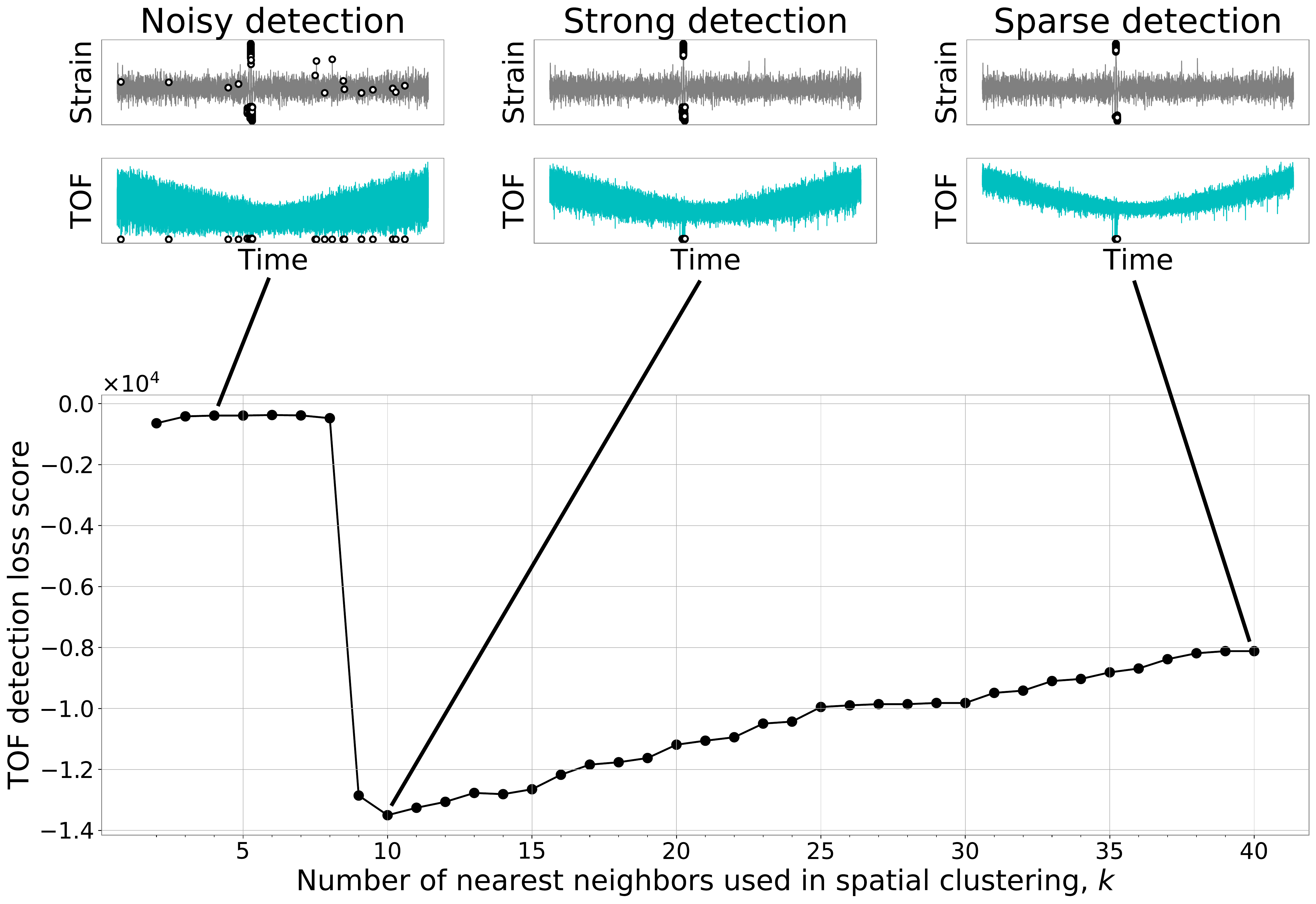}
    \caption{Visual demonstration of how the loss function quantitatively defines a strong anomaly detection. For any particular tunable parameter, there is a range where the detections are noisy, i.e. many points outside of the actual glitch are classified as anomalous. Conversely, there is typically a range where detections are sparse or non-existent. Somewhere in between is a regime where there are many detections at the site of the glitch and few elsewhere. This corresponds to a strong detection and a dip in the loss. In this example the parameter being adjusted is the number of neighbors in TOF's $k$NN clustering step; all three TOF runs shown above are on the same time series.}
    \label{fig:loss_function}
\end{figure}

Thanks to LigoDV-web \cite{ligodv_web}, an extensive database of glitches previously encountered and classified by Gravity Spy is at our disposal. Each of the glitch events in the database is labeled with a time-stamp, providing us with a convenient metric for gauging the accuracy of a TOF detection of that glitch. Given a time series containing a known glitch at time $\tilde{t}$, we run TOF on the time series using a configuration of adjustable parameters $\mathbf{p}$. The output array of anomalous times $\mathbf{a}$ (Equation \ref{eq:tof_anomaly}) can then be quantitatively assessed using the following loss function:
\begin{equation}\label{eq:loss_function}
    \texttt{loss}(\mathbf{p}) = \left[\sum_{t^*\in \mathbf{a}} \frac{\left|t^* - \tilde{t}\right|}{|\mathbf{a}|}\right] - \left[\frac{|\mathbf{a}|}{\sigma(\mathbf{a})}\right]
\end{equation}
where $|\mathbf{a}|$ is the number of elements in $\mathbf{a}$ and $\sigma(\mathbf{a})$ is the standard deviation of $\mathbf{a}$. This loss function is designed to punish detection error while rewarding detections that are dense and tightly clustered. Qualitatively, we characterize a detection as \textit{noisy} if there are many detected anomalies far from the actual glitch time and \textit{sparse} if $\mathbf{a}$ consists of significantly fewer points than can be attributed to the full glitch. In a \textit{strong} detection, $\mathbf{a}$ consists of many data points close to the true glitch time and ideally none elsewhere. As per the convention in machine learning, a lower loss score corresponds to better performance (Figure \ref{fig:loss_function}). A small modification allows $\tilde{t} = [\texttt{t\_start},\ \texttt{t\_end}]$ to be a time range instead of a single timestamp (we simply set all detections in $\mathbf{a}$ that fall within $\tilde{t}$ to have $t^* - \tilde{t} = 0$); this version of the loss function is used in the analyses in Section 4.

\begin{figure}[t]
    \centering
    \includegraphics[scale=0.5]{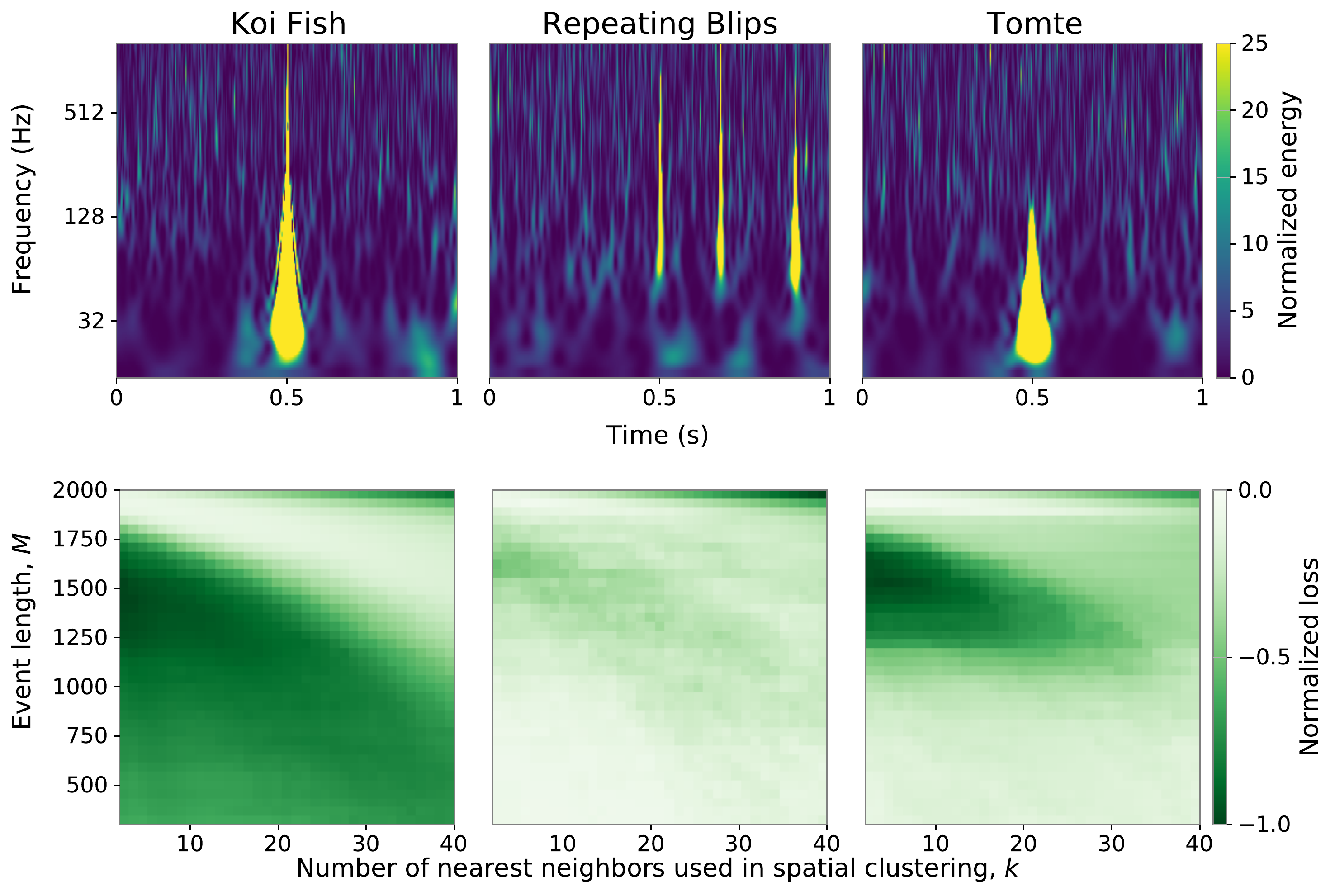}
    \caption{Top plots show representative examples of Koi Fish, Repeating Blips, and Tomte glitches, displayed as Q-transform spectrograms. Bottom plots show the aggregated results of a TOF parameter grid search using 150 examples of each glitch type as evaluated by Equation \ref{eq:loss_function}. Notably, the relative decrease in the loss is significantly less for repeating blips than the other two classes; this is likely due to the multiple individual events causing the overall ``unicorn" to be more loosely clustered overall. We found the parameters for optimal TOF detection characteristics to be $k=9$ and $M=1450$ when considering all three glitch classes. There is another region of low loss at around $k=40$ and $M=1450$, but this parameter configuration tended to capture only one of the sub-events in the repeating blips glitch class (which was undesirable).}
    \label{fig:loss_heatmap}
\end{figure}

To find the optimal configuration of TOF's adjustable parameters, we opted to focus on three known glitch classes occurring on a time scale of less than 0.25 seconds: Koi Fish, Repeating Blips, and Tomtes \cite{detchar}. These morphologies were selected because they share a similar time scale and morphology to potential gravitational wave signals (chirps). We fetched the first 150 examples of each glitch type with classification confidence of over 0.9 listed in LigoDV-web to use as training data. We then performed a grid search over the parameters $k$ and $M$ to find the optimal values (Figure \ref{fig:loss_heatmap}). The time delay was kept at $\tau = 1$, for reasons that will be discussed below. The embedding dimension was set at $d=3$, as higher embedding dimensions incurred heavy computational overhead during the spatial clustering step for no noticeable gain in TOF performance.

\begin{figure}[t]
    \centering
    \includegraphics[scale=0.32]{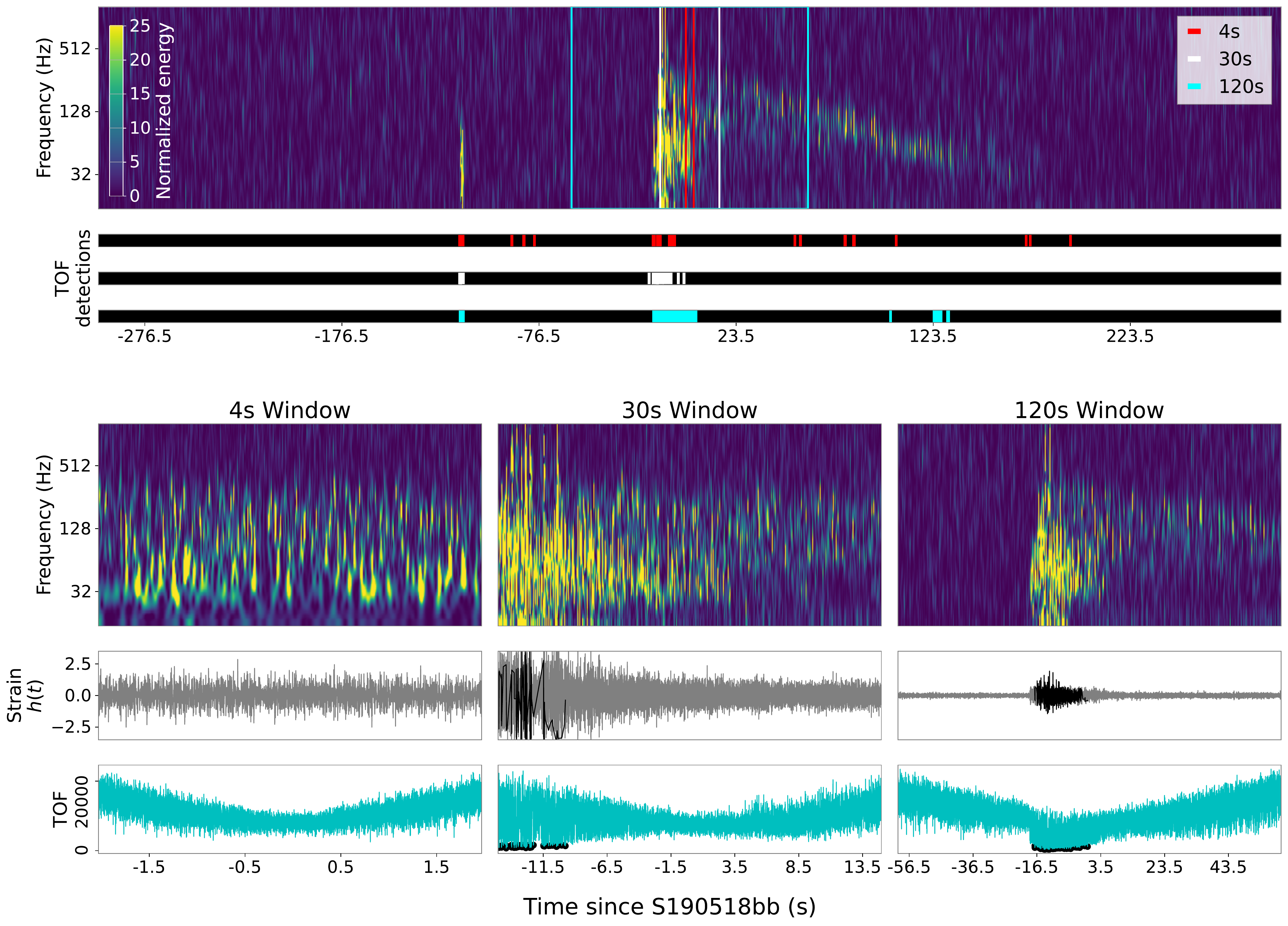}
    \caption{Demonstration of the effect of using windows of different lengths on the TOF algorithm's ability to detect a long noise transient. The spectrogram shown in the top plot is of retracted Open Public Alert S190518bb \cite{eq_gcn}, caused by a nearby earthquake at the LIGO-Hanford interferometer. The bottom plots show TOF's performance over 4, 30, and 120 second windows centered at the event GPS time (with resampling applied to the time-series), shown correspondingly in the top plot as red, white, and blue boxes. Black regions in the bottom plots correspond to TOF detections. TOF is unable to determine the excess energy from the earthquake as an anomaly using the 4-second window. The 30-second window results in a sparse detection. The 120-second window results in a strong detection of the earthquake.}
    \label{fig:resampling}
\end{figure}

We conjecture that most noise transients can be made to look\footnote{At least, for the purposes of the TOF algorithm.} like one of the short glitch classes in Figure \ref{fig:loss_heatmap} through an appropriate linear rescaling of the time axis, as even extremely long noise transients can be made morphologically similar to a short glitch if the time axis is shrunk enough. Thus, we hypothesize it is possible to use a single configuration of TOF parameters to detect anomalies at all scales, provided the TOF algorithm is used in conjunction with a temporal rescaling procedure that windows the data appropriately (Figure \ref{fig:resampling}). Because our implementation of TOF uses a count of data samples as its only measure of lengths of time, temporal rescaling can be implemented via resampling. For this reason, $\tau$ can simply be set to 1, since the content of neighboring data samples is adjusted by resampling. To take advantage of this, we incorporate a \textit{multi-window} scheme that simultaneously searches for noise transients at different scales via the TOF algorithm.

The full anomaly detection pipeline processes detector data in three steps:
\begin{enumerate}[1.]
    \item Raw data for a time interval is whitened and bandpassed between 20 and 500 Hz.
    \item The data is segmented into windows of varying lengths provided by the user, with a base length corresponding to the window size that the TOF parameters were trained on (4 seconds at a sampling rate of 4192 Hz).
    \item TOF runs on each of these windows and attempts to detect unicorns within each window. The results are aggregated into a total anomaly report over the entire time interval of interest.
\end{enumerate}
This is UniMAP. When provided a wide enough range of window sizes, UniMAP is equipped to detect anomalies of arbitrary scale in raw gravitational wave detector data. Of particular note is TOF's ability to detect anomalies of lengths and relative positions within the data window that differ from the anomalies it was trained on. This flexibility allows for the use of significantly fewer windows than time series discord, which must naively attempt every possible window length to cover the same span of possible glitch lengths. The runtime of UniMAP with a fixed number of window lengths is asymptotically linear ($O(n)$ compared to the $O(n^2)$ of time series discord): each individual run of TOF is a constant-time operation due to resampling, and the number of window segments is linear in the length of data. Furthermore the runtime improves drastically (by a constant factor) when using large window sizes, since this decreases the number of windows.

\section{Detecting anomalies in the third observing run}

The Gravitational-Wave Candidate Event Database (GraceDB) \cite{gracedb} lists 80 GW candidates corresponding to Open Public Alerts (OPAs) \cite{abbott2019} issued during LIGO-Virgo's third observing run, between April 1, 2019, and March 27, 2020 (O3). Of these, a subset of 24 are labeled by GraceDB as ``retracted", meaning the event that generated the alert was later found to be non-astrophysical. Additional analysis is performed offline to form a definitive catalog of confidently detected gravitational wave events during the observing run \cite{gwtc2}\cite{gwtc3}. OPAs present opportunities to field-test the TOF pipeline's ability to detect anomalies in longer stretches of strain data. OPAs labeled as retracted in GraceDB (as well as any OPAs later not identified as astrophysical by offline analysis) are of special interest because they correspond to the presence of non-astrophysical noise transients.

GraceDB provides GPS merger times for each of the 80 events listed under O3. The time interval selected for analysis by UniMAP spanned $\pm 150$ seconds around the merger time for each event. Results outside of the sub-interval of $\pm 100$ seconds were truncated to prevent edge effects due to data whitening and bandpassing from possibly causing erroneous TOF detections. The TOF detections on the remaining data segments for each event were then plotted via Q-transform using the same settings as those used for Gravity Spy's image classifier \cite{gspy}.

\begin{figure}
    \centering
    \includegraphics[scale=0.175]{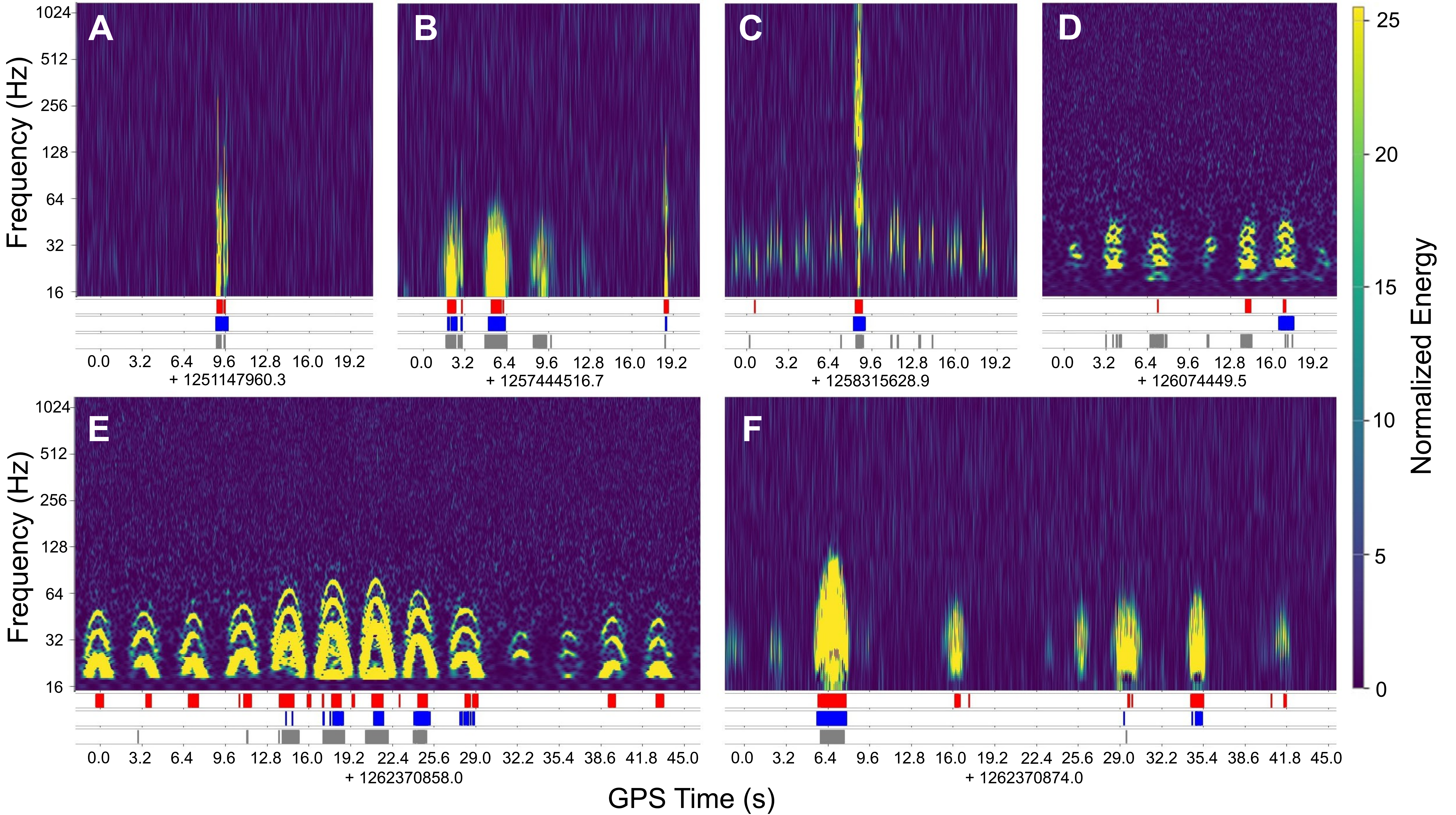}
    \caption{Examples of different glitch morphologies detected by UniMAP during time intervals around OPAs in O3. Red, blue, and grey regions below the Q-transforms correspond to TOF detections from 4, 30, and 120-second windows respectively. The duration of detected glitches spans from around 1 second (A) to well over 40 seconds (E), demonstrating the efficacy of the multi-window resampling method at detecting glitches at different scales.}
    \label{fig:o3_glitches}
\end{figure}

A perusal of the specific detections made by UniMAP on these data segments (Figure \ref{fig:o3_glitches}) shows large diversity in the durations and morphologies of the detected anomalies. This demonstrates the utility of the window-resampling scheme for extending the applicability of the TOF algorithm to arbitrary scales. Of note is the TOF algorithm's existing ability to distinguish between anomalous and non-anomalous regions in unicorns comprised of multiple discrete components (such as Repeating Blips); this allows TOF to detect clusters of periodic glitches when used on long window sizes, as demonstrated in Figure \ref{fig:o3_glitches}D. UniMAP is also able to detect longer periodic glitches, such as the scattering arches shown in Figure \ref{fig:o3_glitches}E. This suggests that combining TOF detections from multiple windows allows the pipeline to decompose periodic noise features into single ``unicorns" within the scope of each window that the basic TOF algorithm can then detect.

\begin{figure}[t]
    \centering
    \includegraphics[scale=0.31]{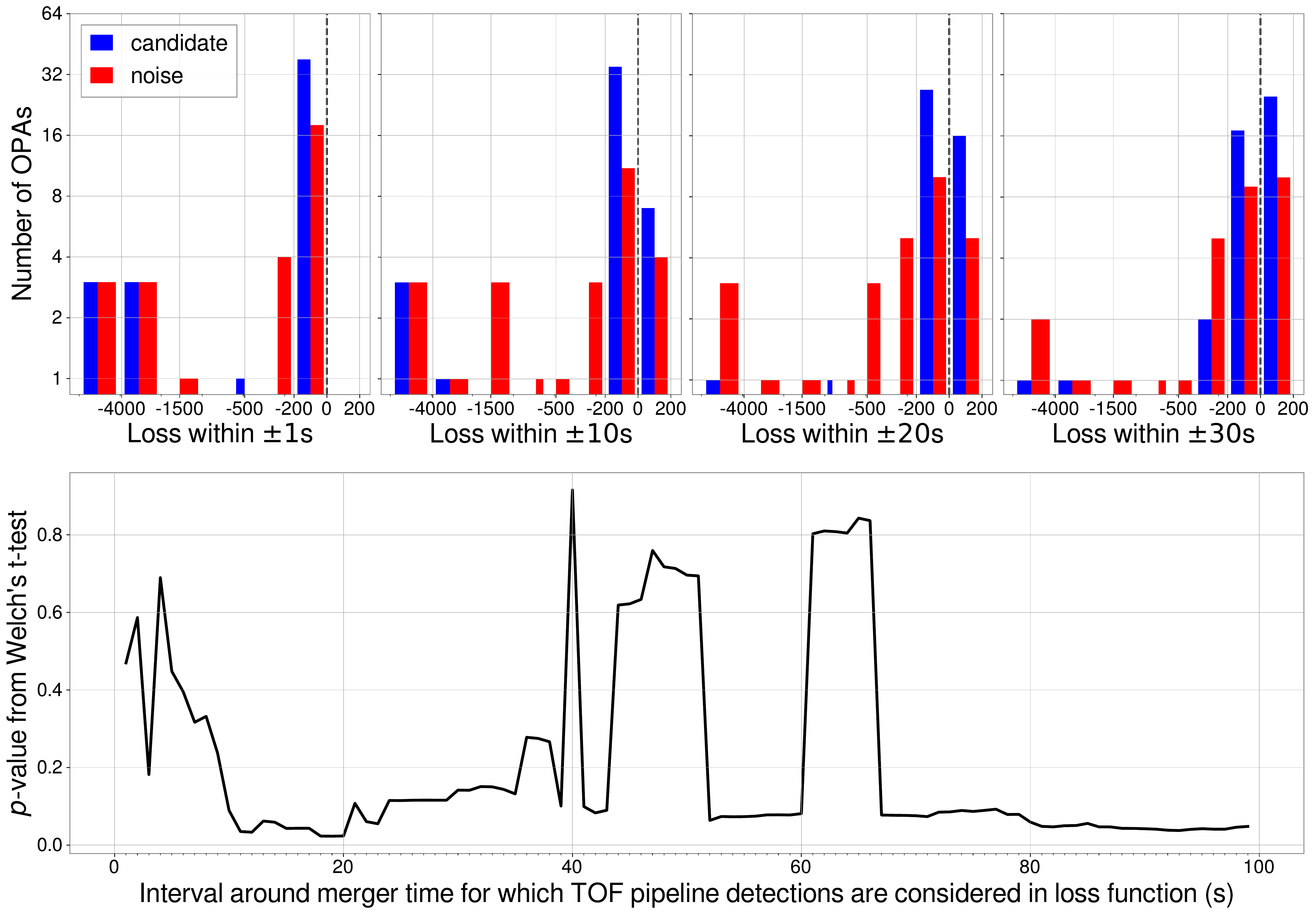}
    \caption{Summary of TOF detection quality near each of the OPAs in O3. For each OPA, we selected the strain data from the interferometer with the highest SNR reading during the recorded event (LIGO-Hanford, LIGO-Livingston, or Virgo) to represent the OPA in the analysis. We analyzed UniMAP detections within a window between $\pm 1$ and $\pm 99$ seconds around the merger time as a quantitative assessment of the TOF detections corresponding to each event. Histograms of the loss scores are shown for windows of length 1, 10, 20 (best performance), and 30 seconds. The bottom plot displays the $p$-value obtained by Welch's t-test against the null hypothesis that detections on the candidate and noise populations have the same mean loss; regions of lower $p$ correspond to window lengths that result in high discrimination between the detection quality of UniMAP on candidate versus noise events as assessed by Equation \ref{eq:loss_function}.}
    \label{fig:opa_losses}
\end{figure}

The second part of the analysis involved a high-level investigation of UniMAP detections near OPAs of different categories. We wanted to determine if UniMAP is more sensitive toward noise events than candidate (astrophysical) events, as we want to avoid the pipeline triggering on true astrophysical signals if possible. Avoidance of flagging real events was not included as part of the TOF parameter optimization, so if candidate events do currently avoid detection by UniMAP, the most likely explanation is that real gravitational wave events do not have high enough signal-to-noise ratios (SNRs) to trigger a TOF detection at current interferometer sensitivities. However, as interferometer sensitivity improves, we may see more gravitational wave events caught by UniMAP; the specific SNR threshold for this is a topic of current research efforts.

For our analysis, we treated all OPAs that were not listed as candidates in GWTC-2.1 and GWTC-3 as noise. Additionally, we excluded events labeled as marginal by the aforementioned GWTCs to reduce the number of categories to two. Due to the relatively small sample size of the remaining events, we do not claim any statistical conclusions from the results presented in Figure \ref{fig:opa_losses}; however, the $p$-values from the t-tests could be used as a rough estimate of the time interval around an event where the discrepancy between TOF detections on a candidate OPA versus a noise OPA is largest. Based on this analysis, this interval seems to be somewhere between 10-20 seconds around the merger time (a full data table of the event losses in the 20-second window is included in Appendix 1). We conclude that in O3, on average, there are features in the strain data in the vicinity of noise events that trigger increased UniMAP detections compared to in the vicinity of true astrophysical events. This may be evidence that detectors experience a higher glitch rate near glitches easily mistaken for astrophysical signals than near true astrophysical signals. Alternatively, these glitches may generally have a wider temporal extent than true astrophysical signals. Confirmation of either of these hypotheses requires more informative means of quantifying UniMAP’s performance near known events than Equation \ref{eq:loss_function} can currently provide. However, these results demonstrate not only the pipeline’s ability to detect glitches of diverse morphologies, but also its potential to help discriminate between noise and astrophysical events.

\section{Conclusions and future work}

The ever-increasing sensitivity of the Advanced LIGO and Advanced Virgo gravitational wave detectors presents the challenge of identifying and filtering noise transients with unexpected morphologies that are difficult to capture via existing model-based methods. The Temporal Outlier Factor provides a model-free, \textit{a priori} means of detecting aperiodic anomalies in time-series data. Used in conjunction with a sliding-window time-resampling data-preprocessing scheme, we have demonstrated that the resulting anomaly detection pipeline is capable of detecting noise transients of widely varying morphologies and durations. Furthermore, a preliminary survey of candidate and noise events in LIGO-Virgo's third observing run suggests that noise events may share common long-duration features that tend to trigger increased TOF detections in their vicinity compared to astrophysical events. Identification of these features may aid in discrimination between true and false positives in future event detection.

We have demonstrated the promising performance of UniMAP; however, there are many directions for further refinement and optimization. The sensitivity of the pipeline to astrophysical signals must be investigated to assess the risk of TOF flagging real events as noise, especially as detector sensitivity increases. Additionally, we require a better quantitative measure for the quality of UniMAP detections on known events than the loss function presented in Equation \ref{eq:loss_function}, particularly one more suited to analyzing short events embedded in long stretches of background. The current static data windowing scheme employed by UniMAP can also be improved via a dynamic windowing scheme that targets regions where anomalies are more likely as assessed either by an alternative anomaly detection algorithm, or recursively by the existing pipeline. Finally, more radical pre-processing schemes for the detector data may make lower-SNR signals more visible to the TOF algorithm. Candidates for such schemes include removal of stationary noise background via existing methods, and conversion of raw strain data into a measure of fractal dimension \cite{marcofractal} before analysis by UniMAP.

In conclusion, UniMAP is a promising new addition to LIGO-Virgo's data quality toolbox and has the potential to be a valuable component in a future glitch detection pipeline; however, further investigation is needed to determine, refine, and optimize its capabilities. We provide the current UniMAP codebase on a public GitHub repository\footnote{\url{https://github.com/JulianZDing/UniMAP}} for anyone interested in trying it for themselves.

\section*{References}
\bibliography{references}

\begin{thebibliography}{10}

\bibitem{aligo}
J~Aasi, B~P Abbott, R~Abbott, et~al.
\newblock Advanced {LIGO}.
\newblock {\em Classical and Quantum Gravity}, 32(7):074001, Mar 2015.

\bibitem{virgo}
F~Acernese, M~Agathos, K~Agatsuma, et~al.
\newblock {Advanced Virgo: a second-generation interferometric gravitational
  wave detector}.
\newblock {\em Classical and Quantum Gravity}, 32(2):024001, dec 2014.

\bibitem{gwtc1}
B.~P. Abbott, R.~Abbott, T.~D. Abbott, et~al.
\newblock {GWTC-1: A Gravitational-Wave Transient Catalog of Compact Binary
  Mergers Observed by LIGO and Virgo during the First and Second Observing
  Runs}.
\newblock {\em Phys. Rev. X}, 9:031040, Sep 2019.

\bibitem{gwtc2}
The LIGO~Scientific Collaboration, the Virgo~Collaboration, R.~Abbott, et~al.
\newblock {GWTC-2.1: Deep Extended Catalog of Compact Binary Coalescences
  Observed by LIGO and Virgo During the First Half of the Third Observing Run},
  2021.

\bibitem{gwtc3}
The LIGO~Scientific Collaboration, the Virgo~Collaboration, the
  KAGRA~Collaboration, et~al.
\newblock {GWTC-3: Compact Binary Coalescences Observed by LIGO and Virgo
  During the Second Part of the Third Observing Run}, 2021.

\bibitem{detchar}
D~Davis, J~S Areeda, B~K Berger, et~al.
\newblock {LIGO detector characterization in the second and third observing
  runs}.
\newblock {\em Classical and Quantum Gravity}, 38(13):135014, Jun 2021.

\bibitem{davis2020}
Derek Davis, Laurel~V White, and Peter~R Saulson.
\newblock Utilizing {aLIGO} glitch classifications to validate
  gravitational-wave candidates.
\newblock 37(14):145001, jun 2020.

\bibitem{noise_sources}
Beverly~K. Berger.
\newblock {Identification and mitigation of Advanced {LIGO} noise sources}.
\newblock {\em Journal of Physics: Conference Series}, 957:012004, feb 2018.

\bibitem{omicron}
Florent Robinet, Nicolas Arnaud, Nicolas Leroy, et~al.
\newblock {Omicron: A tool to characterize transient noise in
  gravitational-wave detectors}.
\newblock {\em SoftwareX}, 12:100620, 2020.

\bibitem{snax}
GstLAL developers.
\newblock {SNAX documentation}, 2020.

\bibitem{Smith2011}
Joshua~R Smith, Thomas Abbott, Eiichi Hirose, et~al.
\newblock {A hierarchical method for vetoing noise transients in
  gravitational-wave detectors}.
\newblock 28(23):235005, nov 2011.

\bibitem{Essick2021}
Reed Essick, Patrick Godwin, Chad Hanna, et~al.
\newblock {iDQ}: Statistical inference of non-gaussian noise with auxiliary
  degrees of freedom in gravitational-wave detectors.
\newblock 2(1):015004, dec 2020.

\bibitem{gspy}
M~Zevin, S~Coughlin, S~Bahaadini, et~al.
\newblock {Gravity Spy: integrating advanced LIGO detector characterization,
  machine learning, and citizen science}.
\newblock {\em Classical and Quantum Gravity}, 34(6):064003, Feb 2017.

\bibitem{Razzano2018}
Massimiliano Razzano and Elena Cuoco.
\newblock Image-based deep learning for classification of noise transients in
  gravitational wave detectors.
\newblock 35(9):095016, apr 2018.

\bibitem{Cuoco2018}
{Wavelet-based classification of transient signals for gravitational wave
  detectors}.
\newblock {\em 26th European Signal Processing Conference (EUSIPCO)}, page
  2648–2652, 2018.

\bibitem{Godwin2020}
{Incorporation of Statistical Data Quality Information into the GstLAL Search
  Analysis}.
\newblock {\em Preprint arXiv 2010.1528}, 2020.

\bibitem{lowlatency}
B.~P. Abbott, R.~Abbott, T.~D. Abbott, et~al.
\newblock {Low-latency Gravitational-wave Alerts for Multimessenger Astronomy
  during the Second Advanced {LIGO} and Virgo Observing Run}.
\newblock 875(2):161, apr 2019.

\bibitem{eq_gcn}
LIGO~Scientific Collaboration and Virgo Collaboration.
\newblock {GCN 24590}.
\newblock \url{https://gcn.gsfc.nasa.gov/other/GW190518bb.gcn3}, May 2019.

\bibitem{eq_retraction}
K.~Kawabe, B.~O'Reilly, and A.~Rocchi.
\newblock {The Retraction of S190518bb}.
\newblock {LIGO Document Control Center}, May 2019.

\bibitem{distancebasedoutlier}
Fabrizio Angiulli and Fabio Fassetti.
\newblock {Distance-based outlier queries in data streams: The novel task and
  algorithms}.
\newblock {\em Data Min. Knowl. Discov.}, 20:290--324, 03 2010.

\bibitem{conformalknn}
Vladislav Ishimtsev, Alexander Bernstein, Evgeny Burnaev, and Ivan Nazarov.
\newblock Conformal $k$-{NN} anomaly detector for univariate data streams.
\newblock In Alex Gammerman, Vladimir Vovk, Zhiyuan Luo, and Harris
  Papadopoulos, editors, {\em Proceedings of the Sixth Workshop on Conformal
  and Probabilistic Prediction and Applications}, volume~60 of {\em Proceedings
  of Machine Learning Research}, pages 213--227. PMLR, 13--16 Jun 2017.

\bibitem{lof}
Markus Breunig, Hans-Peter Kriegel, Raymond Ng, and Joerg Sander.
\newblock {LOF: Identifying Density-Based Local Outliers}.
\newblock volume~29, pages 93--104, 06 2000.

\bibitem{carrera2016ecg}
Diego Carrera, Beatrice Rossi, Daniele Zambon, et~al.
\newblock {ECG Monitoring in Wearable Devices by Sparse Models}.
\newblock In {\em Joint European Conference on Machine Learning and Knowledge
  Discovery in Databases. Springer International Publishing.}, 2016.

\bibitem{hotsax}
E.~Keogh, J.~Lin, and A.~Fu.
\newblock {HOT SAX: efficiently finding the most unusual time series
  subsequence}.
\newblock In {\em Fifth IEEE International Conference on Data Mining
  (ICDM'05)}, pages 8 pp.--, 2005.

\bibitem{tof}
Zsigmond Benkő, Tamás Bábel, and Zoltán Somogyvári.
\newblock {How to find a unicorn: a novel model-free, unsupervised anomaly
  detection method for time series}, 2021.

\bibitem{Takens}
Floris Takens.
\newblock Detecting strange attractors in turbulence.
\newblock In David Rand and Lai-Sang Young, editors, {\em Dynamical Systems and
  Turbulence, Warwick 1980}, pages 366--381, Berlin, Heidelberg, 1981. Springer
  Berlin Heidelberg.

\bibitem{chatterji2004}
S~Chatterji, L~Blackburn, G~Martin, and E~Katsavounidis.
\newblock Multiresolution techniques for the detection of gravitational-wave
  bursts.
\newblock {\em Classical and Quantum Gravity}, 21(20):S1809--S1818, sep 2004.

\bibitem{ligodv_web}
J.S. Areeda, J.R. Smith, A.P. Lundgren, et~al.
\newblock {LigoDV-web: Providing easy, secure and universal access to a large
  distributed scientific data store for the LIGO scientific collaboration}.
\newblock {\em Astronomy and Computing}, 18:27--34, 2017.

\bibitem{gracedb}
LIGO~Scientific Collaboration, Virgo Collaboration, et~al.
\newblock {GraceDB—Gravitational-Wave Candidate Event Database}, 2020.

\bibitem{abbott2019}
B.~P. Abbott, R.~Abbott, T.~D. Abbott, et~al.
\newblock {Low-latency Gravitational-wave Alerts for Multimessenger Astronomy
  during the Second Advanced LIGO and Virgo Observing Run}.
\newblock {\em The Astrophysical Journal}, 875(2):161, Apr 2019.

\bibitem{marcofractal}
Marco Cavaglia.
\newblock Characterization of gravitational-wave detector data with fractal
  analysis.
\newblock {g2net WG3 training school on Machine Learning for Advanced Control
  Techniques}, Aug 2021.

\end{thebibliography}

\newpage
\appendix
\section*{Appendix 1: O3 OPA loss scores (in ascending order)}

Data table for generating the 20-second window histogram in Figure \ref{fig:o3_glitches}. Superevent ID corresponds to the label in GraceDB. The preferred detector corresponds to the detector with the highest SNR for the preferred event in each superevent; this is labeled (unknown) if the SNR data is not available in GraceDB. Each OPA\footnote{The table only contains 77 out of 80 OPAs in O3; three OPAs were later classified as marginal candidates~\cite{gwtc3}.} is categorized as either Candidate or Noise as described in Section 4; candidates labeled with an asterisk (*) represent events that underwent noise-mitigation methods prior to source-parameter estimation in the event catalogs (Table V in GWTC-2 \cite{gwtc2} and Table XIV in GWTC-3 \cite{gwtc3}). The loss was calculated using the merger time as the ``true" time of each event. If no preferred detector was available, the reported loss is the mean loss over all available detectors among H1, L1, and V1.
\begin{center}
\begin{longtable}{ |c|c|c|c|c|c| }
\hline
\multicolumn{6}{|c|}{20 second window} \\
\hline
Row & Superevent ID & Preferred detector & Category & $p$ Terrestrial & Loss\\
\hline
1 & S190808ae & L1 & Noise & 5.732E-01 & -8367.2 \\
2 & S191110af & (Unknown) & Noise & (Unknown) & -6508.7 \\
3 & S190928c & (Unknown) & Noise & (Unknown) & -4997.4 \\
4 & S200112r & L1 & Candidate & 3.375E-04 & -4825.7 \\
5 & S190816i & L1 & Noise & 1.669E-01 & -4178.8 \\
6 & S200308e & L1 & Noise & 1.702E-01 & -3749.2 \\
7 & S191120at & L1 & Noise & 1.717E-01 & -1084.6 \\
8 & S191215w & L1 & Candidate & 2.756E-03 & -980.8 \\
9 & S190829u & L1 & Noise & 1.031E-01 & -616.0 \\
10 & S191117j & L1 & Noise & 1.081E-10 & -502.4 \\
11 & S190524q & L1 & Noise & 7.059E-01 & -442.7 \\
12 & S200108v & L1 & Noise & 1.574E-06 & -437.5 \\
13 & S191213ai & L1 & Noise & 1.524E-01 & -417.7 \\
14 & S200106au & L1 & Noise & 9.280E-01 & -374.3 \\
15 & S200106av & L1 & Noise & 9.962E-01 & -331.9 \\
16 & S191213g & L1 & Noise & 2.320E-01 & -327.6 \\
17 & S190518bb & H1 & Noise & 2.456E-01 & -303.9 \\
18 & S191120aj & L1 & Noise & 3.904E-01 & -177.6 \\
19 & S191110x & L1 & Noise & 7.958E-04 & -142.0 \\
20 & S191109d & L1 & Candidate* & 2.168E-06 & -108.7 \\
21 & S191222n & H1 & Candidate & 3.844E-05 & -58.4 \\
22 & S200311bg & H1 & Candidate & 3.958E-17 & -49.3 \\
23 & S191212q & H1 & Noise & 5.120E-01 & -39.4 \\
24 & S200129m & L1 & Candidate* & 2.001E-24 & -37.5 \\
25 & S191205ah & L1 & Noise & 6.790E-02 & -33.0 \\
26 & S190425z & L1 & Candidate* & 5.974E-04 & -26.0 \\
27 & S190706ai & H1 & Candidate & 1.022E-02 & -23.9 \\
28 & S200116ah & L1 & Noise & 6.579E-05 & -23.5 \\
29 & S190924h & L1 & Candidate* & 4.745E-11 & -19.0 \\
30 & S190510g & L1 & Noise & 5.795E-01 & -15.6 \\
31 & S190521r & L1 & Candidate & 6.677E-04 & -13.0 \\
32 & S190521g & L1 & Candidate & 3.412E-02 & -3.8 \\
33 & S191225aq & L1 & Noise & 6.095E-01 & -2.6 \\
34 & S200224ca & H1 & Candidate & 3.385E-05 & -2.3 \\
35 & S191216ap & H1 & Candidate & 8.428E-16 & -1.8 \\
36 & S200114f & (Unknown) & Noise & (Unknown) & -0.4 \\
37 & S190408an & L1 & Candidate & 9.824E-12 & 0 \\
38 & S190412m & L1 & Candidate & 1.741E-20 & 0 \\
39 & S190503bf & H1 & Candidate* & 1.248E-04 & 0 \\
40 & S190513bm & H1 & Candidate* & 5.999E-08 & 0 \\
41 & S190602aq & L1 & Candidate & 9.665E-03 & 0 \\
42 & S190630ag & L1 & Candidate & 1.792E-07 & 0 \\
43 & S190707q & L1 & Candidate & 1.110E-05 & 0 \\
44 & S190727h & H1 & Candidate* & 4.782E-02 & 0 \\
45 & S190828l & H1 & Candidate & 4.078E-04 & 0 \\
46 & S190915ak & H1 & Candidate & 5.276E-03 & 0 \\
47 & S191105e & L1 & Candidate* & 4.687E-02 & 0 \\
48 & S191129u & L1 & Candidate & 1.163E-27 & 0 \\
49 & S200105ae & L1 & Noise* & 9.727E-01 & 0 \\
50 & S200115j & L1 & Candidate* & 2.893E-04 & 0 \\
51 & S200128d & H1 & Candidate & 3.101E-02 & 0 \\
52 & S200208q & L1 & Candidate & 6.643E-03 & 0 \\
53 & S200213t & H1 & Noise & 3.705E-01 & 0 \\
54 & S200225q & H1 & Candidate & 4.338E-02 & 0 \\
55 & S200302c & H1 & Candidate & 1.104E-01 & 0 \\
56 & S200316bj & L1 & Candidate & 4.263E-03 & 0 \\
57 & S190405ar & H1 & Noise & 9.999E-01 & 0 \\
58 & S190701ah & L1 & Candidate* & 6.563E-02 & 1.2 \\
59 & S190512at & L1 & Candidate & 1.012E-02 & 1.9 \\
60 & S200303ba & V1 & Noise & 1.362E-01 & 3.7 \\
61 & S191124be & V1 & Noise & 7.174E-02 & 3.9 \\
62 & S190930t & L1 & Candidate & 2.574E-01 & 3.9 \\
63 & S191204r & L1 & Candidate & 8.693E-18 & 4.4 \\
64 & S190421ar & H1 & Candidate & 3.260E-02 & 4.5 \\
65 & S190814bv & L1 & Candidate* & (Unknown) & 4.5 \\
66 & S190517h & L1 & Candidate & 4.289E-05 & 4.5 \\
67 & S190519bj & L1 & Candidate & 4.418E-02 & 4.9 \\
68 & S190910d & H1 & Candidate & 2.410E-02 & 5.0 \\
69 & S200219ac & L1 & Candidate & 3.605E-02 & 6.2 \\
70 & S190828j & L1 & Candidate & 3.756E-14 & 6.9 \\
71 & S190718y & H1 & Noise & 9.793E-01 & 8.9 \\
72 & S190923y & L1 & Noise & 3.222E-01 & 9.7 \\
73 & S190901ap & L1 & Noise & 1.393E-01 & 10.2 \\
74 & S190728q & L1 & Candidate & 3.631E-13 & 10.8 \\
75 & S190930s & L1 & Candidate & 4.924E-02 & 11.4 \\
76 & S190426c & H1 & Candidate & 5.750E-01 & 15.3 \\
77 & S190720a & L1 & Candidate & 1.075E-02 & 15.7 \\
\hline
\end{longtable}
\end{center}
\end{document}